**Main Manuscript for**

**Emergent chirality in a polar meron to skyrmion phase transition.**

Yu-Tsun Shao[1], Sujit Das[2,3], Zijian Hong[4,5], Ruijuan Xu[6,7,8], Swathi Chandrika[1], Fernando Gómez-Ortiz[9], Pablo García-Fernández[9], Long-Qing Chen[4], Harold Y. Hwang[6,7], Javier Junquera[9], Lane W. Martin[3,10], Ramamoorthy Ramesh[3,10,11] & David A. Muller[1,12].

[1]Department of Applied and Engineering Physics, Cornell University, Ithaca, New York, USA.
[2]Department of Materials Science and Engineering, University of California, Berkeley, CA, USA.
[3]Materials Research Centre, Indian Institute of Science, Bangalore, India.
[4]Materials Research Institute and Department of Materials Science and Engineering, The Pennsylvania State University, State University Park, PA, USA
[5]Laboratory of Dielectric Materials, School of Materials Science and Engineering, Zhejiang University, Hangzhou 310027, China
[6]Department of Applied Physics, Stanford University, Stanford, CA, USA.
[7]Stanford Institute for Materials and Energy Sciences, SLAC National Accelerator Laboratory, Menlo Park, CAs, USA.
[8] Department of Materials Science and Engineering, North Carolina State University, Raleigh, NC, USA.
[9]Departamento de Ciencias de la Tierra y Física de la Materia Condensada, Universidad de Cantabria, Cantabria Campus Internacional, Avenida de los Castros s/n, 39005 Santander, Spain
[10]Materials Sciences Division, Lawrence Berkeley National Laboratory, Berkeley, CA, USA.
[11]Department of Physics, University of California, Berkeley, CA, USA.
[12]Kavli Institute at Cornell for Nanoscale Science, Ithaca, New York, USA.

*To whom correspondence may be addressed. **Email:** david.a.muller@cornell.edu

**Author Contributions:** Y.-T.S., S.D., R.X., H.Y.H., R.R., and D.A.M. designed research; Y.-T.S., S.D., Z.H., R.X., S.C., F. G.-O., and P.G.-F., and J.J. performed research; Y.-T.S., S.D., and R.X. contributed new reagents/analytic tools; Y.-T.S., S.D., Z.H., F.G.-O, P.G.-F., L.-Q.C., H.Y.H., J.J., L.W.M., and R.R. analyzed data; and Y.-T.S. wrote the paper.

**Competing Interest Statement:** The authors declare no conflict of interest.

**This Word file includes:**

> Main Text
> Figures 1 to 5

**Abstract**

Polar skyrmions are predicted to emerge from the interplay of elastic, electrostatic and gradient energies, in contrast to the key role of the anti-symmetric Dzyalozhinskii-Moriya interaction in magnetic skyrmions. With the discovery of topologically-stable polar skyrmions, it is of both fundamental and practical interest to understand the microscopic nature and the possibility of temperature- and strain-driven phase transitions in ensembles of such polar skyrmions. Here, we explore the reversible transition from a skyrmion state (topological charge of -1) to a two-dimensional, tetratic lattice of merons (with topological charge of -1/2) upon varying the temperature and elastic boundary conditions in $[(PbTiO_3)_{16}/(SrTiO_3)_{16}]_8$ lifted-off membranes. This topological phase transition is accompanied by a change in chirality, from zero-net chirality (in meronic phase) to net-handedness (in skyrmionic phase). To map these changes microscopically required developing new imaging methods. We show how scanning convergent beam electron diffraction provides a robust measure of the local polarization simultaneously with the strain state at sub-nm resolution, while also directly mapping the chirality of each skyrmion. Using this, we demonstrate strain as a crucial order parameter to drive isotropic-to-anisotropic structural transitions of chiral polar skyrmions to non-chiral merons, validated with X-ray reciprocal space mapping and theoretical phase-field simulations. These results revealed by our new measurement methods provide the first illustration of systematic control of rich variety of topological dipole textures by altering the mechanical boundary conditions, which may offer a promising way to control their functionalities in ferroelectric nanodevices using the local and spatial distribution of chirality and order.

**Introduction**

A structural transition in materials involves the rearrangement of a periodic array of motifs in response to external stimuli, which governs the materials' functional properties. Such a transition is not limited to atoms, but also anticipated in lattices consisting of unconventional quasi-particles such as skyrmions or merons[1–4]. The recent discovery of non-trivial topological textures, including flux-closure[5–8], vortices[9–15], skyrmion[16–21], merons[22–24], or hopfions[25] in ferroelectric-oxide nanostructures provides a framework for exploring topology and exotic physical phenomena in condensed matter physics with a focus on polar order[26–28]. Polar skyrmion bubbles consist of three-dimensional (3D) electric dipole textures which, plane by plane, are characterized by an integer topological charge, called the skyrmion number, and defined as

$$N_{\text{sk}} = \frac{1}{4\pi} \iint d^2 r \vec{n} \cdot \left( \frac{\partial \vec{n}}{\partial x} \times \frac{\partial \vec{n}}{\partial y} \right),$$

where $\vec{n}$ is the normalized local dipole moment. Merons are another kind of topological solitons. To understand the differences in their topology, Fig. 1 sketches the dipolar textures for a Bloch-like skyrmion versus a meron. As shown in Figs. 1A-B, at the central *xy*-plane, a polar skyrmion exhibits an out-of-plane polarization ($P_{op}$) at the core and an antiparallel $P_{op}$ outside the boundary (magenta arrows, Fig. 1A-B). The polarization continuously rotates in such a way that the in-plane component shows a curling Bloch-like pattern[29] (blue/yellow arrows, Fig.1A-B) that provides chirality to the skyrmion[20]. In contrast, merons (Figs. 1C-E) exhibit $P_{op}$ at the core and gradually evolve to in-plane $P_{ip}$ at the periphery, without being fully surrounded by an antiparallel $P_{op}$ component[30]. As an example, a meron and a skyrmion can have the same $P_{ip}$ texture but vary only by their $P_{op}$ components. Therefore, simultaneous three-dimensional experimental characterization is necessary to correctly determine the topology of the polar textures, and we show how both in-plane and out-plane polar order can be mapped simultaneously.

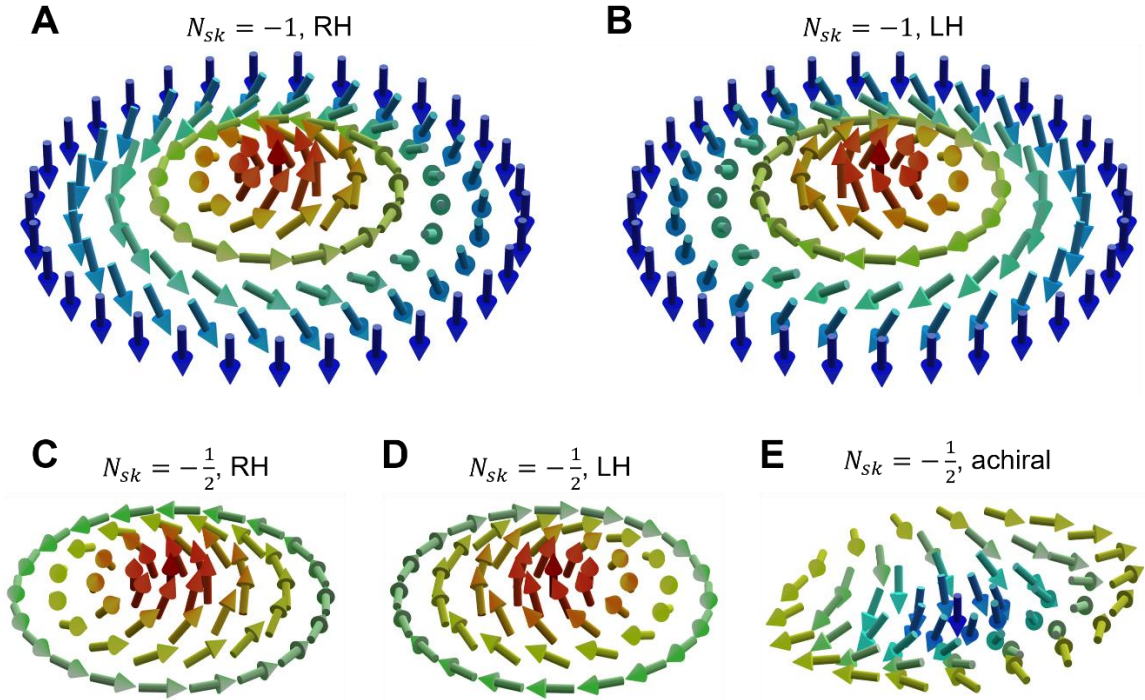

**Figure 1. Sketches of the dipolar textures observed for merons versus skyrmions.** Schematic of Bloch-like skyrmions with (**A**) right-handed and (**B**) left-handed chiralities, both possessing a topological charge of $N_{Sk} = -1$. A common feature of both the Bloch ($\nabla \times \boldsymbol{P} \neq 0, N_{Sk} = -1$) and Néel ($\nabla \cdot \boldsymbol{P} \neq 0, N_{Sk} = -1$) skyrmions is the out-of-plane polarization rotates from maximal at the core, to maximal in the opposite direction at the periphery. Instead, the meronic phases can be characterized by a vanishing out-of-plane polarization at the periphery, and a non-zero value at the core. The in-plane component of the polarization presents a non-vanishing vorticity. In the case of a vortex, both right-handed (**C**), and left-handed (**D**) chiralities are possible, with a topological charge of $N_{Sk} = -1/2$. In the case of an antivortex, and although the total topological charge might remain invariant if we reverse the direction of the out-of-plane polarization (**E**), the final meron configuration is achiral.

A fundamental question pertaining to the spatial arrangement of the skyrmions is the degree of long-range order, if any, amongst the skyrmions, in terms of both the orientational order as well as translational order[4,31]. The emergence of long range order, or conversely the disappearance of a possible long range order in the polar skyrmion lattice driven by temperature can provide a heretofore unexplored possibility of a melting phase transition in such a topologically protected two-dimensional (2D) array of polar skyrmions (akin to the well-known Kosterlitz-Thouless-Halperin-Nelson-Young (KTHNY) transition[4,32]. Although phase field models had predicted the possibility of forming a long-range ordered skyrmion lattice for certain values of mismatch strain with the substrate[33], experimentally this is yet to be demonstrated. Given that the magnitude and sign (compressive vs. tensile) of the strain in the superlattice is a critical component of such long-range order, we sought to manipulate this by lifting off the superlattice from the substrate[34–36]. In doing so, we are able to study the ground state of the skyrmions without any interference from substrate constraint. On such a free-standing membrane, we then imposed different elastic boundary conditions by varying the temperature, due to differences in thermal expansion of the two materials, to manipulate the degree of long-range order in the skyrmion lattice.

Here, we report the direct observation of sequential structural transformations for the polar skyrmions: from stripe-shaped to circular-shaped disordered skyrmions bubbles, to a tetratic-ordered meron lattice in lifted-off $[(PbTiO_3)_{16}/(SrTiO_3)_{16}]_8$ superlattice membranes through integrated experimental measurements and theoretical phase-field calculations. Electron imaging of the polar distortions can be challenging due to local sample mis-tilt artifacts which can affect or dominate the contrast. To disentangle tilt and polarization, we developed an approach for the analysis of Kikuchi bands by using the intensity asymmetry in diffuse scattering, recorded by four-dimensional scanning transmission electron microscopy (4D-STEM) with an electron microscopy pixel array detector (EMPAD)[37]. Using the high-dynamic-range EMPAD, we were able to collect the full scattering distribution, including Kikuchi bands and Bragg reflections (>$10^4$ difference in diffracted intensities), which also contains lattice information for determining the strain state of the specimen. Using 4D-STEM and dynamical diffraction analysis we also demonstrate, for the first time, a direct experimental determination of chirality of individual polar skyrmions. Our observations reveal the emergence of a square meron lattice with $N_{sk}$=-1/2 from the disordered skyrmion phase with a $N_{sk}$=-1, in which the chirality changed from left-handed to zero-net chirality, respectively.

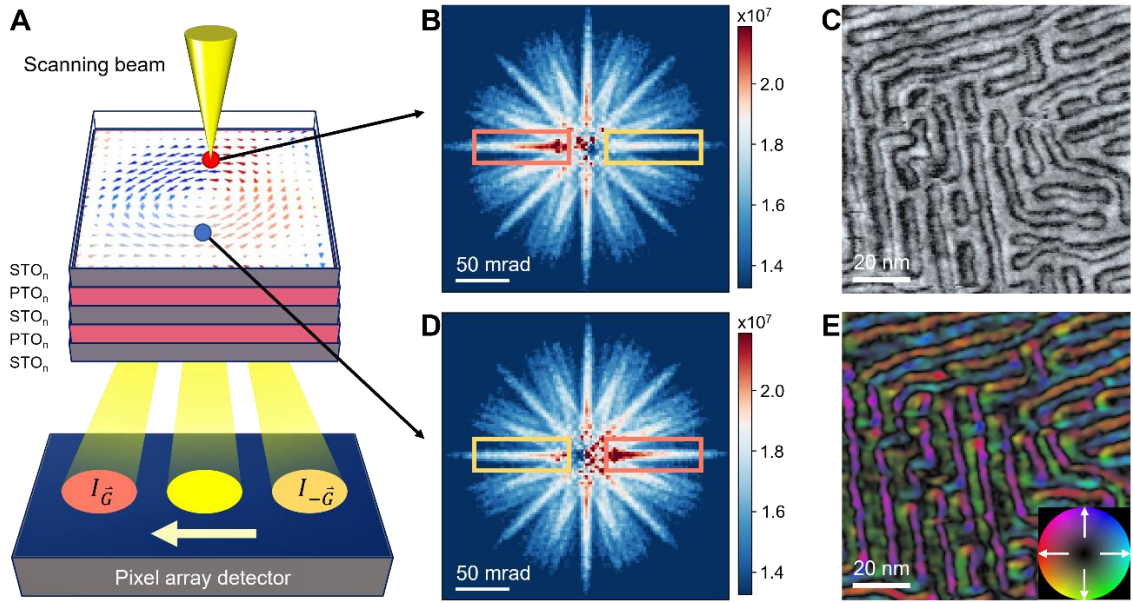

**Figure 2. Imaging in-plane polarization textures.** (**A**) Schematic of the plan-view 4D-STEM imaging technique on the [(PbTiO$_3$)$_{16}$/(SrTiO$_3$)$_{16}$]$_8$ superlattice which uses a scanning electron probe and pixelated array detector, where a diffraction pattern was recorded at each probe position. The local polarization direction can be determined by observing the difference of diffracted intensities of Friedel pairs, $I_{\vec{G}}$ and $I_{-\vec{G}}$. Representative diffraction patterns taken from (**B**) top and (**D**) bottom of a skyrmion, where the polarity-sensitive Kikuchi bands in the thermal diffuse scattering are selected for determining polarization, as marked by pink and yellow boxes. For clarity of display, the Kikuchi bands intensity were weighted by k$^2$, where $\vec{k}$ denotes the scattering vector from the transmitted spot. (**C**) Plan-view dark field STEM image of a (SrTiO$_3$)$_{16}$/(PbTiO$_3$)$_{16}$/(SrTiO$_3$)$_{16}$ trilayer reconstructed from the 4D-STEM dataset showing nanometer-size round and elongated features. (**E**) Polarization map reconstructed using Kikuchi

bands from the same region as **(C)** showing the in-plane Bloch components of polar-skyrmions. The color map represents the in-plane polarization direction at each point.

## Results

**Experimental setup of 4D-STEM.**

To examine the local polarization distribution, we performed 4D-STEM experiments on plan-view, lifted-off samples (Fig. S1) to image the in-plane Bloch components (details in Methods). Briefly, 4D-STEM works by using an EMPAD detector which records the 2D electron diffraction pattern over a 2D grid of probe positions, resulting in 4D datasets (Fig. 2A, details in Methods)[38–40]. As a result of dynamical diffraction effects, the charge redistribution associated with ferroelectric polarization leads to the breakdown of Friedel's law[41,42]. Thus, due to channeling effects (Fig. S2), the polarization field within the top $PbTiO_3$ layer can be measured quantitatively from intensity differences of polarity-sensitive Kikuchi bands[43,44] (Fig. 2B & 2D) or Bragg reflections[45,46] (Fig. S3). We note that due to different channeling conditions, for plan-view imaging, atomic-resolution STEM is more sensitive to the top of the skyrmion, where the Néel component dominates, while the 4D-STEM method is more sensitive to the middle of the skyrmion where Bloch component dominates. To illustrate these channeling effects, we have performed multislice simulations of HAADF-STEM (semi-convergence angle of 21.4 mrad) and 4D-STEM datasets (semi-convergence angle of 2.45 mrad) on a model polar skyrmion obtained from $2^{nd}$-principles calculations (Fig. S4). We employed these Kikuchi bands for polarity mapping as they are less sensitive to artifacts such as disinclination strain or crystal mis-tilts, which are inevitable in ferroic oxides[47,48] (Fig. S5 & S6). These Kikuchi bands are generally weak features require detectors of high dynamic range which is only available until very recently[37]. The fast detectors also allow for the mapping of polar textures which outruns the charging effects caused by the focused electron beam. For example, Fig. 2C shows a high-angle annular dark-field (HAADF) image of the skyrmions from plan-view reconstructed from the 4D-STEM dataset, where the out-of-plane polarization ($P_{op}$) is separated by domain walls with circular or elongated features. At the same region (same 4D-STEM dataset), Figure 2E shows the in-plane polarization ($P_{ip}$) map of Bloch-like rotation can be reconstructed by using the polarity-sensitive Kikuchi bands. The white arrows and colors in Fig. 2E denote the direction of $P_{ip}$, whereas the saturation represents the vector magnitude. The dark color indicates the $P_{op}$ regions.

**Observation of topological phase transition in polar textures.**

Upon heating from 223 K to 373 K, we observed successive structural transitions in the skyrmion ensemble, from striped to circular-shaped polar skyrmions to a tetratic-ordered lattice. Figure 3A shows the polarization configuration at 223 K, which consists of elongated stripes of ~100 nm in length. In analogy with in-plane cuts ($Q_x - Q_y$) from X-ray reciprocal space maps (RSM), the fast Fourier transform (FFT) patterns of plan-view HAADF images indicate the in-plane ordering of polar textures (Fig. S7). For example, two peaks were found in the FFT pattern along the pseudocubic $\pm(100)_{pc}$ directions (inset, Fig. 3A) consistent with the stripe features.

When heated to 298 K, the stripes deformed into circular shapes (Figs. 3B & 3E) of ~10 nm in diameter. The packing of circular skyrmions appears to be random and isotropic, i.e., no preferred orientational order, as indicated by the halo in the FFT pattern (inset, Fig. 3B). With further increase of temperature to 373 K, the random skyrmion arrangements were replaced by an ordered tetratic arrangement (Fig. 3C), confirmed by the four peaks in $\pm(100)_{pc}$ and $\pm(010)_{pc}$ directions in the corresponding FFT. Figs. 3D-F show the corresponding polarization maps at different temperatures obtained from 4D-STEM datasets. For the stripe- (223 K) and disordered, circular-shaped (298 K) skyrmions, the polarization maps (Figs. 3D & 3E) indicate that the maximum $P_{ip}$ is observed at the periphery of the skyrmions, whereas the minima (almost zero) are at the core and outside the boundaries. Figs. 3G-I show the corresponding phase-field simulations of polar textures at different temperatures and under various strain conditions, in which the average effective in-plane lattice constants are obtained from experimental measurements of >80,000 convergent beam electron diffraction (CBED) patterns (Fig. S8). The phase field simulations demonstrate a systematic change from the labyrinthine skyrmions at low temperature (223K) to an ordered tetratic structure at higher temperatures (373K), driven not by a pure thermal effect but by the changes in the in-plane strain state originating from lattice thermal expansion.

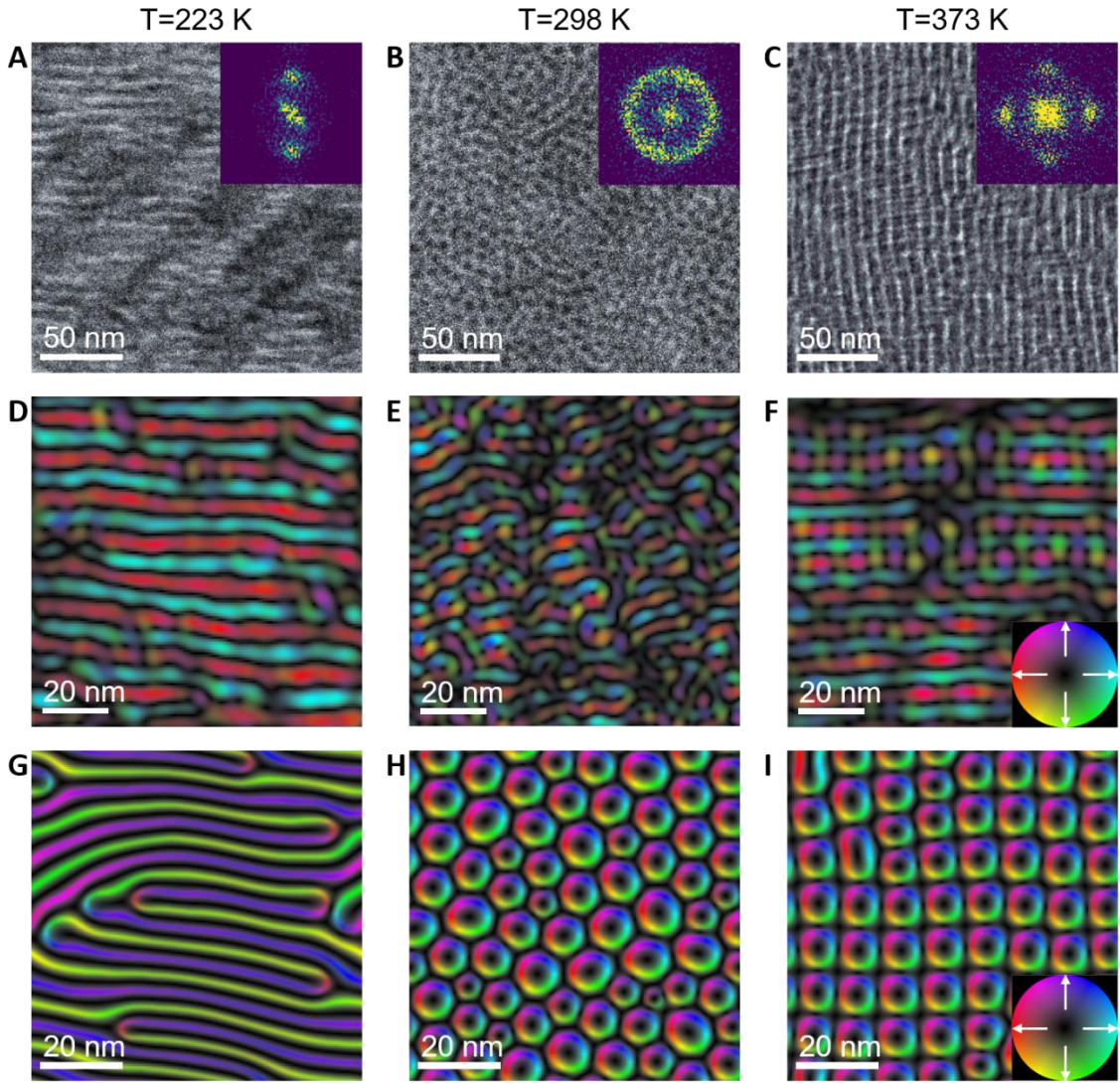

**Figure 3. Variations of polar textures with temperature.** Plan-view dark field STEM imaging of [(SrTiO$_3$)$_{16}$/(PbTiO$_3$)$_{16}$]$_8$ superlattice acquired with temperatures at (**A**) 223 K, (**B**) 298 K, and (**C**) 373 K. Insets, fast Fourier transform (FFT) of the images in A-C showing different types of ordering. Polarization maps reconstructed from the 4D-STEM dataset of superlattice at (**D**) 223 K, (**E**) 298 K, and (**F**) 373 K, showing the in-plane Bloch components of polar-skyrmions. Figures (**D**)-(**F**) are acquired from nearby regions of (**A**)-(**C**). The corresponding phase-field simulations with temperatures and in-plane lattice parameters (*a,b*) of (**G**) 223 K, *a*=3.875 Å, *b*=3.885 Å; (**H**) 298 K, *a*=*b*=3.905 Å; and (**I**) 373 K, *a*=*b*=3.899 Å. The color wheel hue (saturation) corresponds to the direction (magnitude) of the in-plane component of the ferroelectric polarization. We note

that the phase field simulations capture the qualitative changes in morphology of particle-like objects as a function of temperature/strain conditions. However, while the general symmetry of the phase is predicted, the microscopic details of polar vectors at the boundaries require first-principles based calculations.

To compare the difference of polar textures, regions in Fig. 4A-B (yellow box) were selected to show the $P_{ip}$ polarization maps (Figs. 4C-D). The white arrows and colors in Fig. 4C-D denote the magnitude and the direction of $P_{ip}$, whereas the saturation represents the vector magnitude. The dark color indicates the $P_{op}$ regions at skyrmion cores and outside the boundary, which are separated by Bloch domain walls ($P_{ip}$) consistent with cross-section data (Fig. S9). The $P_{op}$ at the skyrmion cores points positively towards the growth direction ([001], or +z) and are antiparallel to $P_{op}$ outside of the boundary, labeled as green dots (+z) and red crosses (-z), respectively (Fig. 4E). The Bloch components ($P_{ip}$) exhibit a continuous rotation of the local polarization vector forming a closed loop, as illustrated in a map of the curl of the polarization vector field $(\nabla \times \mathbf{P})_{[001]}$ (Fig. 4A). Both elongated and circular skyrmions exhibit $P_{ip}$ having clockwise (CW) rotation at the periphery yielding a vorticity of +1. Combining $P_{ip}$ and $P_{op}$ information, we confirm that both elongated and circular skyrmions manifest with a skyrmion number of $N_{sk}$=-1.

The most striking observation is the appearance of an ordered structure at 373 K (Figs. 4D & 4F), which represents a square meron lattice with $N_{sk}$=-1/2. Fig. 4F clearly demonstrates these periodic arrays, which also indicates that the maximum $P_{ip}$ polarization is observed at the periphery, whereas the minimum (almost zero) is at the core. Three types of core regions were observed, which we labeled according to the cores of $P_{ip}$ having CW rotation (green), counterclockwise rotation (CCW; blue), and antivortices (red). The dot in the circle indicates the $P_{op}$ pointed out of the page (along the growth direction), while the cross indicates $P_{op}$ pointing into the page. From a cross-section polarization map (Fig. S9), the $P_{op}$ at the cores of vortices (vorticity of +1) and antivortices (vorticity of -1) are plausibly antiparallel. On the other hand, $P_{op}$ appears at vanishing points of $P_{ip}$, implying the $P_{op}$ at vortex cores are not fully surrounded by $P_{op}$ of the opposite direction (Fig. S10). From this, we can compute the skyrmion number[3]:

$$N_{\text{sk}} = \frac{1}{4\pi} \iint d^2 r \vec{n} \cdot \left( \frac{\partial \vec{n}}{\partial x} \times \frac{\partial \vec{n}}{\partial y} \right) = \frac{1}{2} v \cdot (P_{op}^p - P_{op}^c),$$

where $v$ represents the vorticity, $P_{op}^{c}$ the value of the out of plane polarization at the core and $P_{op}^{p}$ the value of the out of plane polarization at the periphery. We deduce the vortices (blue and green dots) to be merons with a topological number of $N_{Sk} = \frac{1}{2} 1 \cdot (0-1) = -\frac{1}{2}$, and the antivortices as merons[25] with $N_{Sk} = \frac{1}{2}(-1) \cdot (0-(-1)) = -\frac{1}{2}$ (Fig. 1).

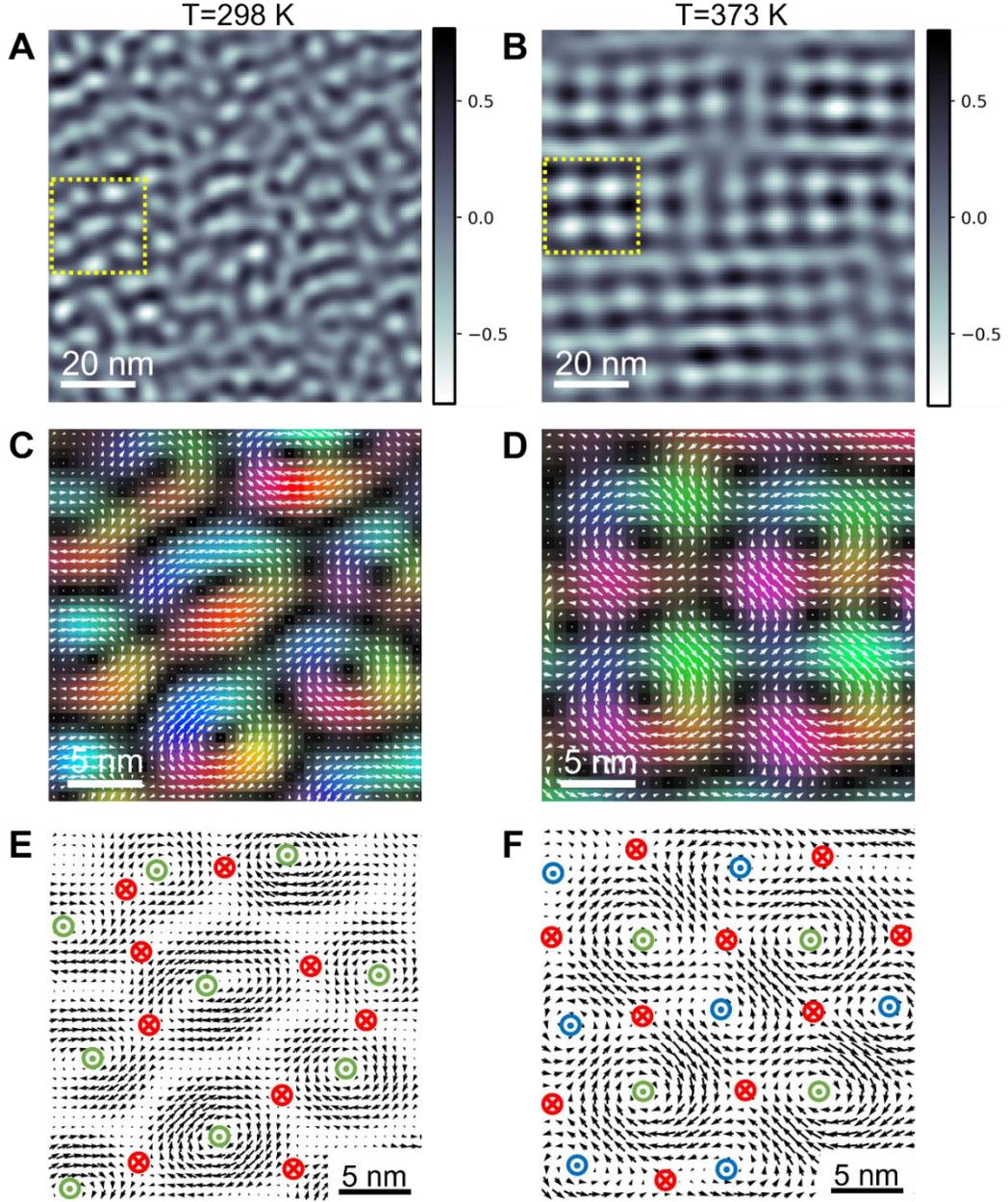

**Figure 4. Real-space observations of disordered polar skyrmions and a square lattice of merons.** The curl of in-plane polarization $(\vec{\nabla} \times \vec{P})_{[001]}$ showing the rotation directions of the structures at temperatures of (**A**) 298 K and (**B**) 373 K. (**C** and **D**) Enlarged in-plane polarization mapping from the yellow box region of A and B, respectively exhibiting the skyrmion texture at 298 K and local ordered meron textures at 373 K. (**E** and **F**) Details of (**C** and **D**), where vortices (clockwise: green, counterclockwise: blue) and antivortices (red) are labeled. The dots in circles represent polarization pointing out of the page, while the cross points into the page.

**Imaging chirality of individual polar texture.**

While 4D-STEM works well for mapping $P_{ip}$ polarizations, we have so far only speculated about the corresponding $P_{op}$ at skyrmion cores based on the cross-section data (Fig. S9). To test our hypothesis, it is necessary to experimentally determine the chirality of the 3D polar-vector field for each skyrmion, which poses challenges for projection techniques such as TEM. Fortunately, we can overcome this problem by utilizing the dynamical diffraction effects in higher-order Laue zone (HOLZ) reflections, which was established to retrieve 3D structural information such as handedness of chiral crystals[49,50]. In this study, we specifically examine the intensity differences of chirality sensitive Bijvoet pairs[51], such as $(671)/(\bar{6}71)$ and $(771)/(\bar{7}71)$ (yellow box, Fig. 5A).

First, we use dynamical diffraction simulation of a right-handed skyrmion and meron as a reference. For a right-handed skyrmion or meron, the diffracted intensity for $(\bar{6}71)$ is stronger than that of $(671)$ at this exact incident beam direction, while other pairs of reflections remain approximately symmetrical (Fig. 5C, black curve). With this in mind, we can determine the chirality of an individual skyrmion by comparing intensity variations of Bijvoet pairs in a 4D-STEM dataset. At 298 K, we carefully selected regions with minimal crystal mis-tilts and determined that both elongated and circular skyrmions are left-handed (Fig. S11). An ordered tetratic lattice appears upon heating to 373 K, as shown in the $(771)/(\bar{7}71)$ intensity ratio map (Fig. 5B). In this region, two types of chiral structures are identified, labeled as #1 (red) and #2 (blue). HOLZ intensity profiles show that $(\bar{6}71)$ is weaker than $(671)$ from the diffraction pattern averaged over meron type #1, and the opposite case for meron type #2 (Fig. 5C, red and blue curves, respectively). Using dynamical diffraction simulations as reference (Fig. 5C, dashed gray curve), we can thus determine the chirality for meron type #1 as left-handed, and type #2 as right-handed. The alternating chirality of types #1 & #2 shown in Fig. 5B is consistent with our previous hypothesis of out-of-plane polarizations shown in Fig. 4F, labeled as green and blue vortices, respectively. More interestingly, the diffraction pattern averaged over a larger field of view (Fig. 5B, black box) shows symmetrical intensity, indicating the loss of chirality (Fig. 5C, black curve). Note that our observation does not violate the Poincaré-Hopf theorem[16,52], where a periodic array of vortices and antivortices yields zero net vorticity, yet imposing no constraints on the chiralities in such topological textures.

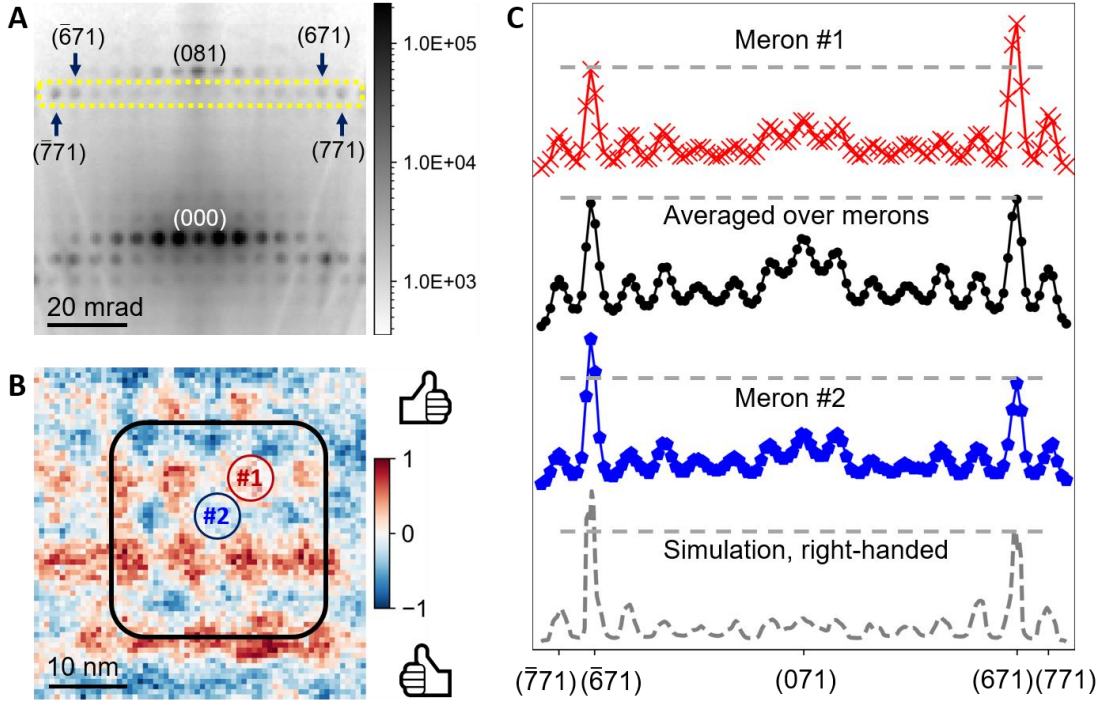

**Figure 5. Measurement of handedness of the chiral polar textures using 4D-STEM.** (**A**) A representative experimental diffraction pattern acquired at 373 K (from the meron square lattice phase), oriented ~6.2° away from [001] zone axis, and tilted along one of the mirror planes. (**B**) Map of normalized intensity difference between (771) and ($\bar{7}71$) reflections reconstructed from the 4D-STEM dataset – i.e. one diffraction pattern is recorded at each spatial pixel in panel B. The positive (negative) regions indicate the polar textures having left-handed (right-handed) chirality. (**C**) Intensity line profiles of HOLZ reflections from diffraction patterns taken in from the selected regions #1, #2 labeled in (**B**), displaying the intensity difference between two pairs of reflections: (671) and ($\bar{6}71$), (771) and ($\bar{7}71$). The colored labels in (**B**) indicates the region from which the diffraction patterns of panel (**C**) were extracted. The colored diffraction intensity profiles with markers show experimental data, whereas the simulation of a right-handed meron is shown in dashed grey line at the bottom. The dashed yellow box in (**A**) shows the region of diffraction space integrated to produce the HOLZ line profiles in panel C.

**Discussion**

The transition from skyrmions to merons has been reported in chiral magnets with increasing in-plane magnetic anisotropy[2,30,53]. In 2D ferroelectrics, from the mesoscopic symmetry-breaking perspective, the observed isotropic-to-anisotropic structural transition upon heating was recently reported as an "inverse transition" of labyrinthine domains[22]. Such a transition was driven by the increase of thermal fluctuations of dipoles, which led to the domain reorientation or coalescence associated with the annihilation of merons and antimerons. In case of polar vortices, thermal fluctuations can lead to the melting of chiral crystals and even the loss of chirality[54]. On the other hand, in the case of polar skyrmions, the origin of the observed transition from skyrmions to a meron lattice can be attributed to changes in the elastic boundary conditions. The strain imposed on the $[(PbTiO_3)_{16}/(SrTiO_3)_{16}]_8$ lifted-off membranes is caused by the thermal stress upon heating/cooling. In our experiment, the oxide membranes are ~4× higher in thermal expansion coefficients than the $SiN_x$ TEM grid to which they are attached[55,56]. The oxide membranes are suspended over a 2 µm hole, the strain is applied from the edge of the hole due to thermal stress, and therefore small locally varying anisotropy within which gives small local bias to the texture orientation. As a consequence, the lifted-off membrane is under compressive strain at 373 K and tensile strain at 223 K, with additional local bending. In case of 4D-STEM experiments, we deliberately searched for flat regions to avoid artifacts in measuring $P_{ip}$. A detailed 4D-STEM analysis of averaged in-plane lattice parameters (>80,000 diffraction patterns) shows that the lifted-off membrane is locally more rectangular at 223 K than at 373 K (Fig. S8). Experimentally, the lattice constant is strongly temperature dependent. To decouple the influence of strain versus purely thermal effects on the phase transition, phase-field simulations were carried out at controlled strain boundary conditions for the range of measured temperatures (Fig. S12). For a fixed in-plane lattice constant, simulation results for temperatures from 223-373K showed a similar disordered skyrmion phase, indicating that here strain plays an important role in the skyrmion ordering. Furthermore, the change in anisotropy of in-plane lattice parameters, as seen in the 4D-STEM strain analysis (inset, Fig. S8D), is also consistent with FFT patterns for stripes and square lattice (Figs. 3A & 3C). Thus, we anticipate the occurrence of a long-range ordered meron lattice at room temperature in $(PbTiO_3)_n/(SrTiO_3)_n$ superlattices grown on a compressive substrate such as $LaAlO_3$.

In summary, we report the observation of a topological phase transition sequence in polar skyrmions, whereby tuning the temperature and strain boundary conditions, the anisotropic stripe phase deforms into an isotropic disordered circular phase, and finally transforms into an anisotropic ordered phase. This is the first observation of such transitions in a dipolar topological structure, in which the skyrmions deform in shape and a square lattice of merons appears, thereby preserving the topological charge of the system. The chiralities for each phase were also determined experimentally at nm-scale, for the first time. We hope that the microscopic observation of such a topological phase transition will stimulate further work to explore the macroscopic manifestation of the changes in topology with strain, which clearly is the critical external stimulus (in contrast to magnetic systems, where the Dzyalozhinskii-Moriya coupling plays a key role). Finally, our findings imply that such dipolar textures are a fertile ground for exploring new phases and topology, and with possible applications in nanoscale ferroelectric logic and storage devices.

## Methods

**Deposition of Thin Film Superlattices and Membrane Lift-Off.**

The epitaxial lift off sacrificial layer of 16 nm $Sr_2CaAl_2O_6$ with a 2.4 nm $SrTiO_3$ capping layer was synthesized on single-crystalline $SrTiO_3$ (001) substrates via reflection high-energy electron diffraction (RHEED) – assisted pulsed laser deposition. Subsequent to this, $n$-$SrTiO_3$/$n$-$PbTiO_3$/$n$-$SrTiO_3$ trilayers ($n$- is the number of monolayers; $n=16$) and $[(PbTiO_3)_{16}/(SrTiO_3)_{16}]_8$ superlattices were synthesized *ex-situ* on this template via RHEED-assisted pulsed-laser deposition (KrF laser). RHEED was used during the deposition to ensure the maintenance of a layer-by-layer growth mode for both the $PbTiO_3$ and $SrTiO_3$. (*Sample Preparation using RHEED-assisted Pulsed-laser Deposition*). The specular RHEED spot was used to monitor the RHEED oscillations. The heterostructure was first spin-coated with a polymer support of 500 nm thick polymethyl methacrylate (PMMA) film and placed in deionized water at room temperature until the sacrificial $Sr_2CaAl_2O_6$ layer was fully dissolved. The PMMA coated film was then released from the substrate and transferred onto the TEM grid. Finally, the PMMA layer was dissolved and removed from the membrane in acetone. Additional details can be found in *Membrane Lift-off and Transfer*.

**4D-STEM for Polarization and Chirality determination.**

We performed 4D-STEM experiments using an electron microscopy pixel array detector (EMPAD), where the 2D electron diffraction pattern was recorded over a 2D grid of real space probe positions, resulting in 4D datasets. Due to dynamical diffraction effects, the charge redistribution associated with polarization leads to the breakdown of Friedel's law (*4D-STEM for Polarization Mapping*). From the collected CBED patterns, the polarization direction in the plan-view samples is reconstructed by calculating the center-of-mass in polarity sensitive Kikuchi bands along the cubic directions. We employ the Kikuchi bands as a more robust means to extract polarity information against internal crystal mis-tilts, which often occurs in ferroic oxides due to disinclination strain. Since chirality is a 3D property and TEM is a projection technique, we can retrieve this 3D information by utilizing the intensity asymmetries of chirality sensitive Bijvoet pairs in higher order Laue zone (HOLZ) reflections. Additional details for chirality determination are discussed in *4D-STEM for Chirality Determination*.

**Phase-field simulations.**

Detailed expressions of the energy terms, materials parameters as well as the numerical simulation procedure is described in ref(10). A periodic boundary condition is used along the in-plane dimensions, while a superposition method is applied in the out-of-plane direction. To determine the local strain state, the reference pseudocubic lattice constants for STO (PTO) at 223K, 300 K and 373 K are set as 3.901 Å (3.953 Å), 3.905 Å (3.957 Å) and 3.909 Å (3.961 Å), respectively. The average effective substrate lattice constants are taken from experimental measurements, which are set as $a$=3.875 Å, $b$=3.885 Å for 223 K, $a$=$b$=3.905 Å for 300 K and $a$=$b$=3.899 Å for 373 K. The large compressive strains are due to local bending when the lift-off membrane is heated/cooled down. We believe that the dominant effect is the global thermal induced large compressive strain under heating, thus, for the simplicity of the model, we ignore the local strain inhomogeneity. A background dielectric constant of 40 is used. Random noise with a magnitude of 0.0001 C/m$^2$ is added to the system as the initial polarization distribution of the system. Additional details of the simulation parameters and the resulting structures are discussed in *Phase-Field Calculations*.

**Acknowledgments**

The authors acknowledge fruitful discussions with Prof. Jian-Min Zuo, Prof. Kin Fai Mak, Dr. Shengwei Jiang, and Zui Tao. Funding was primarily provided by the Department of Defense, Air Force Office of Scientific Research under award FA9550-18-1-0480. The electron microscopy studies were performed at the Cornell Center for Materials Research, a National Science Foundation (NSF) Materials Research Science and Engineering Center (DMR-1719875). The Cornell FEI Titan Themis 300 was acquired through NSF-MRI-1429155, with additional support from Cornell University, the Weill Institute and the Kavli Institute at Cornell. The authors thank M. Thomas, J. G. Grazul, M. Silvestry Ramos, K. Spoth for technical support and careful maintenance of the instruments. The materials synthesis work is supported by the Quantum Materials program from the Office of Basic Energy Sciences, US Department of Energy (DE-AC02-05CH11231). The membrane lift-off techniques were developed with support from US Department of Energy, Office of Basic Energy Sciences, Division of Materials Sciences and Engineering, under contract number DE-AC02-76SF00515. The phase-field simulation work is supported as part of the Computational Materials Sciences Program funded by the U.S. Department of Energy, Office of Science, Basic Energy Sciences, under Award No. DE-SC0020145. F.G.-O., P.G.-F., and J.J. acknowledge financial support from Grant No. PGC2018-096955-B-C41 funded by MCIN/AEI/10.13039/501100011033 and by ERDF "A way of making Europe," by the European Union. F.G.-O. acknowledges financial support from Grant No. FPU18/04661 funded by MCIN/AEI/10.13039/501100011033. R. R. acknowledges support from the Army Research Office under the ETHOS MURI via cooperative agreement W911NF-21-2-0162. L.W.M. acknowledges support from the U.S. Department of Energy, Office of Science, Office of Basic Energy Sciences, under Award Number DE-SC-0012375 for the development and study of ferroic heterostructures.

**Competing interests**

The authors declare no competing interests.

**Data availability**

Experimental and simulation data (4D-STEM dataset and python script for analysis) are provided in the figures and Supplementary Information, and are publicly available at https://doi.org/10.5281/zenodo.74940363. Additional data are available from the corresponding authors upon request.

# Supplementary Information for

# Emergent chirality in a polar meron to skyrmion phase transition.

Yu-Tsun Shao[1], Sujit Das[2], Zijian Hong[3,4], Ruijuan Xu[5,6], Swathi Chandrika[1], Fernando Gómez-Ortiz[7], Pablo García-Fernández[7], Long-Qing Chen[3], Harold Y. Hwang[5,6], Javier Junquera[7], Lane W. Martin[2,8], Ramamoorthy Ramesh[2,8,9] & David A. Muller[1,10].

Corresponding author: David A. Muller

Email: david.a.muller@cornell.edu

**This Word file includes:**

Supplementary text

Figures S1 to S14

SI References

**Supplementary Information Text**

**Supplementary notes on topology of a polar meron.** From a topology perspective, the most important difference between a skyrmion and a meron is the evolution of out-of-plane component of the polarization from the core to periphery. In a skyrmion, the out-of-plane component changes from a direction at the core to the opposite direction at the periphery. Whereas in a meron, the out-of-plane component changes from a direction at the core to be totally confined in-plane at the periphery. As an example, a meron and a skyrmion can have the same in-plane texture and vary only by their out-of-plane components. Take the merons shown in Fig.4 for an example, the core region (with an out-of-plane component of the polarization) is not fully surrounded by a region with the antiparallel out-of-plane polarization. Instead, now each region with an out-of-plane component of the polarization is confined within a region with in-plane polarization at the periphery.

Second, the regions with an opposite component of the out-of-plane polarization *always appear at the center of an anti-vortex* (see the red crosses in Figs. 4B, 4D, 4F, and Fig. S9B). Therefore, the presence of an out-of-plane component of the polarization does not penalize the appearance of the meronic phase. Just the opposite, it enhances it since the regions surrounding the up and down polarizations contribute with the same skyrmion number of -1/2, as shown in Fig. 1.

**Sample preparation using RHEED-assisted pulsed-laser deposition.** The epitaxial lift off sacrificial layer of 16 nm $Sr_2CaAl_2O_6$ with a 2.4 nm $SrTiO_3$ capping layer was synthesized on single-crystalline $SrTiO_3$ (001) substrates via reflection high-energy electron diffraction (RHEED)–assisted pulsed laser deposition. The growth of the $Sr_2CaAl_2O_6$ layer was carried out in a dynamic argon pressure of $4\times10^{-6}$ Torr, at a growth temperature of 710 °C, a laser fluence of 1.35 J/cm$^2$, and a repetition rate of 1 Hz, using a 4.8 mm$^2$ imaged laser spot. The growth of the $SrTiO_3$ layer was conducted in dynamic oxygen pressure of $4 \times 10^{-6}$ torr, at a growth temperature of 710 °C, a laser fluence of 0.9 J/cm$^2$, and a repetition rate of 1 Hz, using a 3.0 mm$^2$ imaged laser spot. The heterostructure was then cooled down to room temperature at the growth pressure. Subsequent to this, $n$-SrTiO$_3$/$n$-PbTiO$_3$/$n$-SrTiO$_3$ trilayers ($n$- is the number of monolayers; $n$=16) and $[(PbTiO_3)_{16}/(SrTiO_3)_{16}]_8$ superlattices were synthesized *ex-situ* on this template via RHEED-assisted pulsed-laser deposition (KrF laser). The PbTiO$_3$ and the SrTiO$_3$ were grown at 610 °C in 100 mTorr oxygen pressure. For all materials, the laser fluence was 1.5 J/cm$^2$ with a repetition rate of 10 Hz. RHEED was used during the deposition to ensure the maintenance of a layer-by-layer growth mode for both the PbTiO$_3$ and SrTiO$_3$. The specular RHEED spot was used to monitor the RHEED oscillations. After deposition, the heterostructures were annealed for 10 minutes in 50 Torr oxygen pressure to promote full oxidation and then cooled down to room temperature at that oxygen pressure.

**Membrane lift-off and transfer.** The heterostructure was first spin-coated with a polymer support of 500 nm thick polymethyl methacrylate (PMMA) film and placed in deionized water at room temperature until the sacrificial $Sr_2CaAl_2O_6$ layer was fully dissolved. The PMMA coated film was then released from the substrate and transferred onto the TEM gird (NH050D2, Norcada Inc.). Finally, the PMMA layer was dissolved and removed from the membrane in acetone, as schematically shown in Fig. S12. The membrane sample surface was further cleaned in ozone at 180 °C for 10 mins.

**X-ray structural analysis.** In order to obtain a comprehensive picture of the crystal structure of superlattices, as well as information on the in-plane and out-of-plane ordering, was carried out using a Panalytical X'Pert Pro X-ray Diffraction (XRD) diffractometer with Cu-$K_\alpha$ radiation ($\lambda$ = 1.5405 Å). The high crystalline quality of the films, the smooth nature of the interfaces and the skyrmion ordering, was confirmed from reciprocal space mapping of the as-grown and lifted-off superlattice.

**Phase-field calculations.** Phase-field simulations were performed to study equilibrium polar structures of $(PTO)_{16}/(STO)_{16}$ superlattice under different temperatures and strain conditions. The spontaneous polarization vector $\vec{P}$ is used as the primary order parameter. The temporal evolution of $\vec{P}$ is governed by the time dependent Ginsburg-Landau equation, i.e.,

$$\frac{\partial P_i}{\partial t} = -L\frac{\delta F}{\delta P_i} \quad (i = 1,3)$$

Where $L$ is the kinetic coefficient, t is the evolution time. The total free energy $F$ can be obtained by integrating the contributions from the mechanical, electrical, Landau chemical and polar gradient energies,

$$F = \int (f_{Elastic} + f_{Electric} + f_{Landau} + f_{Gradient}) dV$$

Detailed expressions of the energy terms, materials parameters as well as the numerical simulation procedure is described in the published literature[1–5]. The simulation system is discretized into a three-dimensional grid of $200\Delta x \times 200\Delta y \times 350\Delta z$, with $\Delta x = \Delta y = \Delta z = 0.4$ nm. A periodic boundary condition is used along the in-plane dimensions, while a superposition method is applied in the out-of-plane direction[6]. In the out-of-plane direction, the thickness of the substrate, film and air are set as $30\Delta z$, $288\Delta z$ and $32\Delta z$, respectively; while the film is comprised of periodic stacking of $16\Delta z$ of PTO layers and $16\Delta z$ of STO layers. A closed-circuit, electric boundary condition is assumed where the electric potential is fixed to zero at the top and bottom of the film surface[2]. When a thin film boundary condition is applied where the stress on the top of the thin film is zero, and the displacement at the bottom of the substrate sufficiently far away from the film/substrate interface is set to zero[3]. An iteration-perturbation method is adopted to account for the inhomogeneity in the elastic constants of PTO and STO[7]. To determine the local strain state, the reference pseudocubic lattice constants for STO (PTO) at 223K, 300 K and 373 K are set as 3.901 Å (3.953 Å), 3.905 Å (3.957 Å) and 3.909 Å (3.961 Å), respectively. The average effective substrate lattice constants are taken from experimental measurements, which are set as $a$=3.875 Å, $b$=3.885 Å for 223 K, $a$=$b$=3.905 Å for 300 K and $a$=$b$=3.899 Å for 373 K. The large compressive strains are due to local bending when the lift-off membrane is heated/cooled down. A background dielectric constant of 40 is used[8,9]. Random noise with a magnitude of 0.0001 C/m² is added to the system as the initial polarization distribution of the system. The polar structures are shown in Figs. 2G-I & S8. At 223K, due to the anisotropic bending of the membrane during cooling, elongated skyrmion stripes, or bimerons, are formed along *X*-axis. Meanwhile, at room temperature, a disordered skyrmion lattice is observed, consistent with a previous report[10]. When the system is further heated up to 373K, a locally ordered square skyrmion lattice is observed. The formation of this locally ordered square lattice is attributed to the large local compressive strain that generated due to the bending of the membrane during heating. Interestingly, some dislocation-like features are also observed, which could locally disturb the ordered structure, consistent with experimental observations.

**STEM.** The plan-view samples of the $[(PbTiO_3)_{16}/(SrTiO_3)_{16}]_8$ superlattices were lifted-off and transferred to TEM grids. Cross-sectional TEM specimens were prepared on the same plan-view samples, using a FEI Strata 400 focused ion beam (FIB) with a final milling step of 2 keV to reduce damage. The initial sample surface was protected from ion-beam damage by depositing carbon and platinum layers prior to milling. The cross-sectional TEM specimen has a thickness of ~25 nm as determined by CBED analysis. HAADF-STEM images were recorded by using a Cs-corrected FEI Titan operated at 300 keV, with beam semi-convergence angle of 21.4 mrad and beam current of 30 pA.

**SCBED for polarization mapping.** We performed scanning convergent beam electron diffraction (SCBED) experiments using an electron microscopy pixel array detector (EMPAD), where the 2D electron diffraction pattern was recorded over a 2D grid of real space probe positions, resulting in 4D datasets. Experimental data was acquired using a FEI Titan operated at 300 keV with 15 pA beam current, 2.45 mrad semi-convergence angle, having a probe of ~8 Å FWHM (full-width at half-maximum). A double-tilt liquid-nitrogen-cooled Gatan specimen holder was used for temperature-dependent studies ranging from 95 K to 373 K. The CBED patterns were captured by the EMPAD with exposure time set to 1 ms per frame, for which a 256 × 256 scan can be recorded in under 2 minutes. Due to dynamical diffraction effects, the charge redistribution associated with polarization leads to the breakdown of Friedel's law. From the collected CBED patterns, the polarization direction in the plan-view samples is reconstructed by calculating the center-of-mass in polarity sensitive Kikuchi bands along the cubic directions. We employ the Kikuchi bands as a more robust means to extract polarity information against internal crystal mis-tilts, which often occurs in ferroic oxides due to disinclination strain. In addition, due to electron channeling effects, the polarity information obtained from Kikuchi bands at this experimental condition is mostly arising from the topmost $PbTiO_3$ layer, which overcomes the problem of overlapping signals from $PbTiO_3$ multilayers projected along the plan-view geometry (Fig. S2). For $SrTiO_3/PbTiO_3/SrTiO_3$ trilayer samples, Friedel pairs of Bragg reflections were used, such as $(300)/(\bar{3}00)$ and $(030)/(0\bar{3}0)$ for x and y components of polarization, respectively (Fig. S3). By matching with dynamical diffraction simulations (Fig. S5), we can unambiguously determine the polarization directions in real space. We note that due to different channeling conditions from different probe semi-convergence angles of 21.4 mrad and 2.45 mrad, for plan-view imaging, atomic-resolution STEM and 4D-STEM map the Néel and Bloch components of a polar skyrmion, respectively (Fig. S4). For clarity of display, the polarization maps obtained using the Kikuchi bands method were first subtracted a low frequency background, then smoothed using a Gaussian filter of 3 pixels. The details in filtering are shown in Fig. S13.

**SCBED for chirality determination.** In order to excite the higher order Laue zone (HOLZ) reflections, the plan-view samples were deliberately tilted ~6.2° away from the [001] zone axis, along one of the mirror planes (Fig. S10A). We then perform SCBED experiments exactly at this diffraction condition at various temperatures. Within a SCBED dataset, we carefully selected local regions with minimal tilt and thickness variations. By using dynamical diffraction simulations as the reference, the chirality can thus be determined by comparing the intensity asymmetry of Bijvoet pairs, such as $(671)/(\bar{6}\bar{7}1)$ and $(771)/(\bar{7}\bar{7}1)$.

**SCBED for strain analysis.** We perform exit wave power cepstrum (EWPC) analysis on SCBED datasets to look at changes in lattice parameters. The EWPC works by a discrete Fourier transform of the logarithm of a CBED pattern, resulting units in real-space. Figure S7B shows an EWPC pattern, in which the peaks correspond to projected Pb-Pb inter-atomic distances. Thus, the change in mean projected, in-plane lattice parameters can be measured by comparing the peak distances in EWPC patterns. Sub-picometer precision can be achieved by sub-pixel peaking fitting using the algorithm described previously[11]. For the sake of self-consistency, SCBED datasets for temperature-dependent strain analysis were acquired using the exact same TEM optics, within one experimental session. To produce the histogram in Fig. S7, each diffraction pattern was probed from a 1nm diameter spot, and strain from that diffraction pattern was calculated. We recorded 128x128=16,384 diffraction patterns from a $100\times100$ nm$^2$ region. The histogram was based on 5 of such regions, which is >80,000 diffraction patterns. Consequently, what we have is a histogram of local in-plane lattice parameter variations over an area of 500 nm$^2$, rather than an averaged lattice parameter. In other words, SCBED is probing the distributions of local textures rather than the ensemble average.

**Dynamical diffraction simulation.** The CBED simulations were carried out using the µSTEM software[12], with neutral atomic scattering factors of Waasmaier & Kirfel[13]. The atomic coordinates were taken from results of 2$^{nd}$-principles simulations for a right-handed polar skyrmion. $25\times25$ diffraction patterns with a 3.2-Å scan step size were simulated at 300-keV beam energy and with 2.45-mrad semi-convergence angle. To simulate Kikuchi bands, thermal diffuse scattering effect was included with the frozen-phonon approximation.

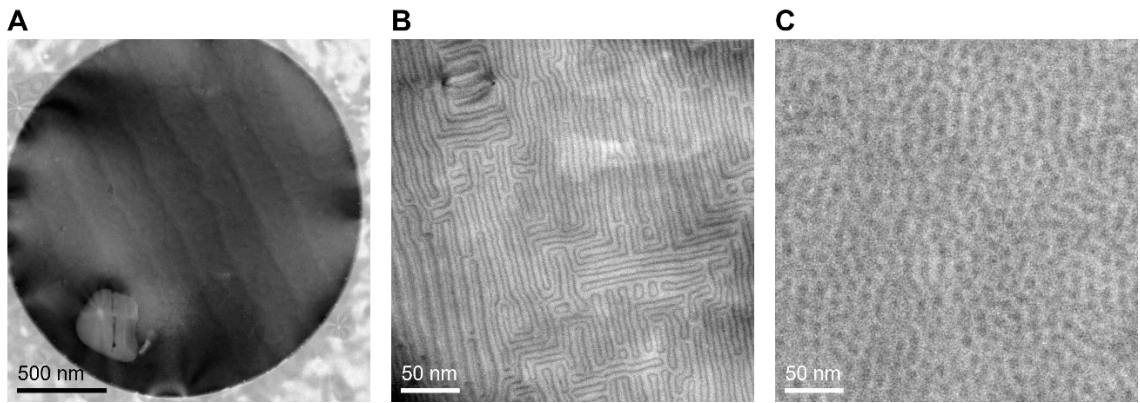

**Fig. S1. Lifted-off, freestanding membrane of $(PbTiO_3)_{16}/(SrTiO_3)_{16}$ heterostructures for plane-view TEM.** (**A**) Low-magnification STEM image showing the oxide membrane suspended over a ~2 µm $SiN_x$ hole of the TEM grid. Exemplar STEM images of (**B**) a trilayer and (**C**) a superlattice acquired from freestanding regions and at room temperature.

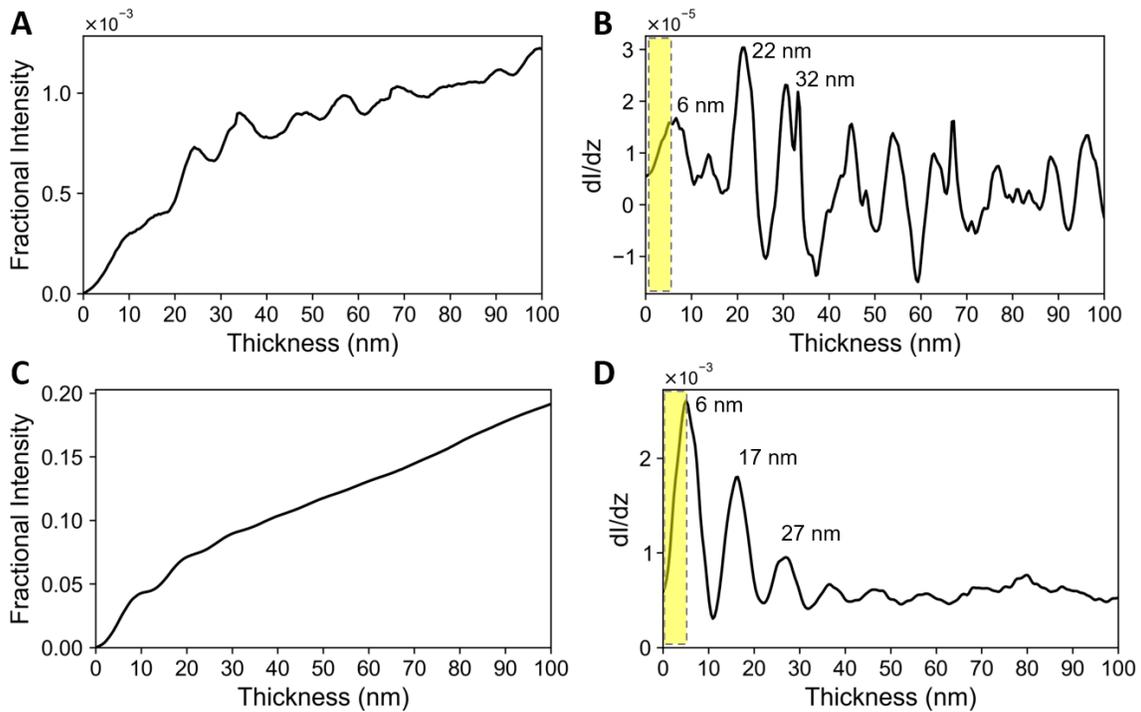

**Fig. S2. Electron beam channeling along a column of Sr atoms in SrTiO₃.** The intensity $I(z)$ of (**A**) (300) Bragg reflection and (**C**) high-angle ADF signal (40-100 mrad) for a 300-keV electron beam centered on the Sr-site as a function of depth $z$ into the crystal. The focused probe with semi-convergence angle of 2.45 mrad is similar to that used in the EMPAD experiments to separate diffraction disks. (**B**) The derivative of the (300) diffracted intensity, showing several peaks of signal are generated at 6 nm, 22 nm, 32 nm into the SrTiO₃, corresponding to the points where 1st,

2nd, and 3rd PbTiO₃ layers would begin in the multilayer structure. (**D**) From derivative $dI/dz$, the signal is channeling most efficiently at 6 nm. The yellow box shows the thickness of 16 unit cells of SrTiO₃. By changing the collection angle, the Kikuchi bands are more suitable than (300) reflections for retrieving polarization information from a single (top) PbTiO₃ layer in the repeated cell in the 16×16 multilayer structure.

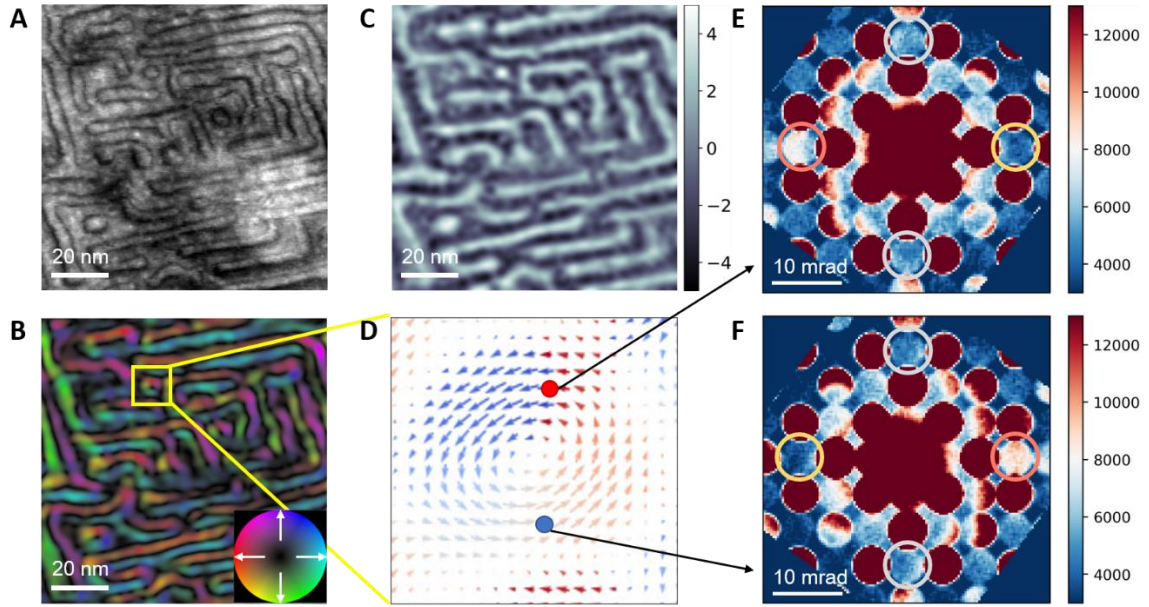

**Fig. S3. Imaging internal Bloch components of polar-skyrmions in the $(SrTiO_3)_{16}/(PbTiO_3)_{16}/(SrTiO_3)_{16}$ trilayer.** (**A**) <100> dark field image and (**B**) polarization map reconstructed from the SCBED dataset showing the in-plane Bloch components of polar-skyrmions and labyrinth phase. The color map in (**B**) represents the in-plane polarization direction at each point. (**C**) Curl of the in-plane polarization. (**D**) Magnified skyrmion from the yellow box in (**B**). Ferroelectric polarization direction can be determined by observing the difference of diffracted intensities of Friedel pairs, $I_{\vec{G}}$ and $I_{-\vec{G}}$. Representative CBED patterns taken from (**E**) top and (**F**) bottom of a skyrmion, where the polarity-sensitive <300> Bragg reflections are selected for determining polarization, as marked by pink and yellow circles.

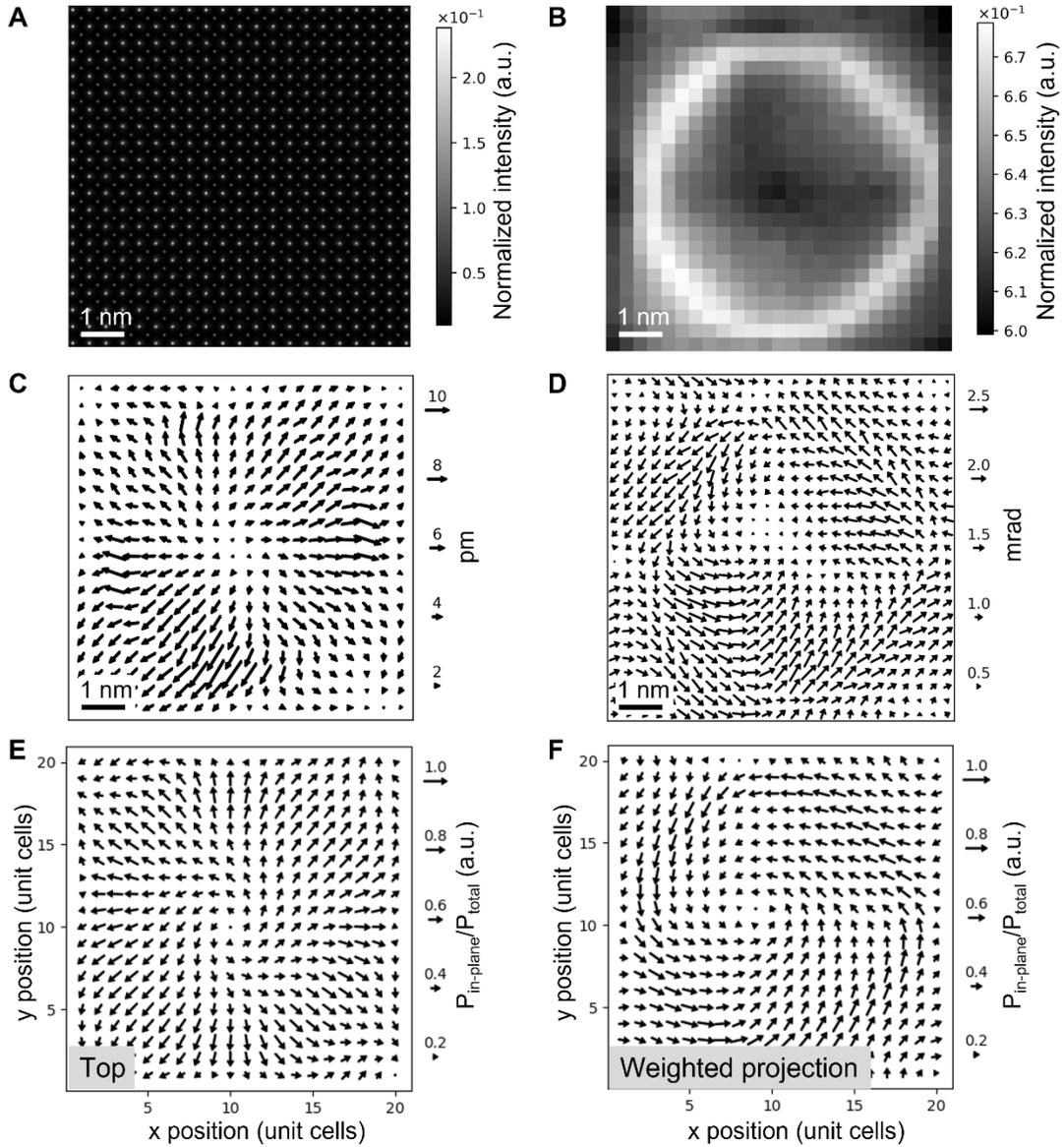

**Fig. S4. Plan-view polarization mapping of a polar skyrmion using atomic-resolution HAADF-STEM versus 4D-STEM Kikuchi bands.** Simulated images for both (**A**) atomic resolution HAADF-STEM and (**B**) ADF-STEM image generated from the multislice-calculated 4D-STEM dataset. (**C**) Ti-displacement vector map obtained from (A), and (D) polarity map obtained from Kikuchi bands in the simulated 4D-STEM dataset in (B). The in-plane polarization component obtained from second-principles calculations of the skyrmion structure showing (**E**) hedgehog-like structure at the top and (**F**) Bloch vortex from weighted projection with more on the central plane in PbTiO$_3$. The HAADF-STEM and 4D-STEM datasets are simulated using a 300 keV electron, with probe semi-convergence angles of 21.4 mrad and 2.45 mrad, respectively. Due to difference in channeling conditions (the more convergent HAADF-STEM dechannels much closer to the entrance surface), HAADF-STEM and 4D-STEM are sensitive to Néel and Bloch components of the polar skyrmion, respectively.

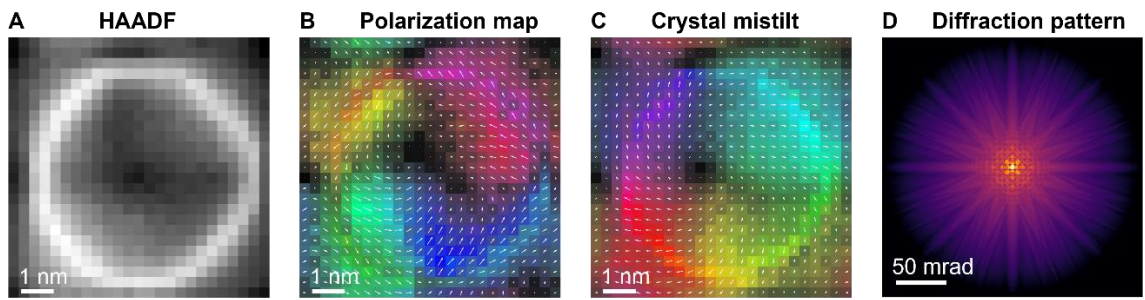

**Fig. S5. Decoupling polarization and crystal mistilts in a polar skyrmion.** The 4D-STEM/SCBED dataset of a polar skyrmion simulated using coordinates from second-principles calculations, with a 300-keV electron beam and a focused probe semi-convergence angle of 2.45 mrad. (**A**) High angle annular dark field (HAADF) image generated by integrating an angular range of 50-130 mrad. (**B**) Polarization map obtained using the Kikuchi band method. (**C**) Effective crystal mistilts information can be extracted from the intensity redistribution in the (000) disk. (**D**) Averaged CBED pattern shown in logarithmic scale. Note the maps of polarization and crystal mistilts can be decoupled using different parts in the CBED patterns.

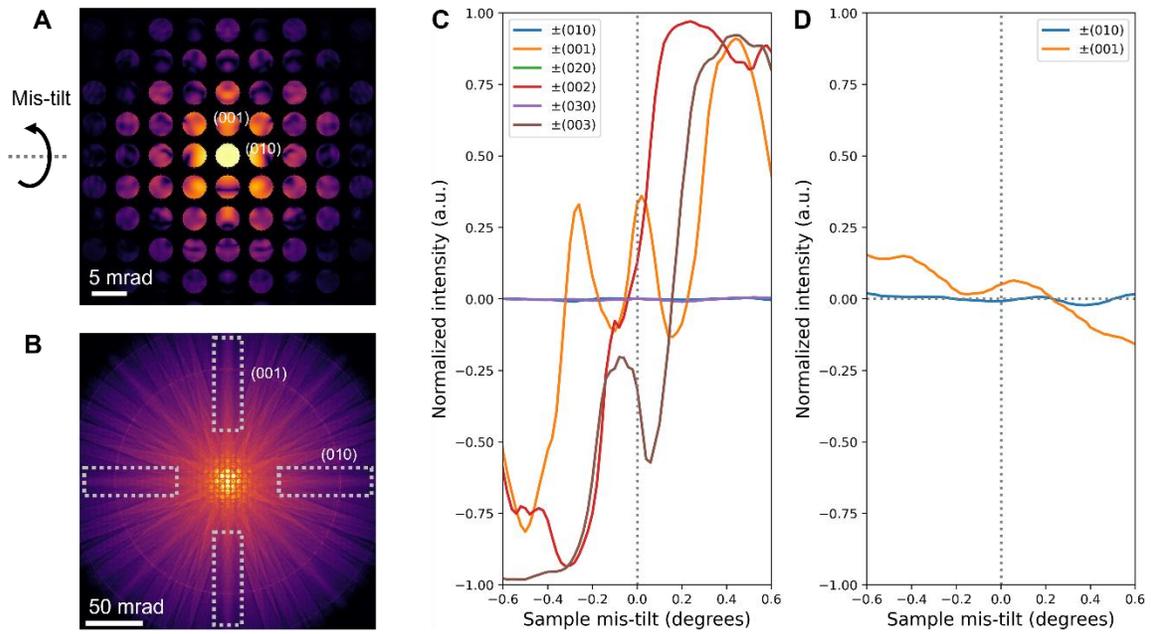

Fig. S6. **Diffraction effects and mis-tilt artifacts on the polarity determination of PbTiO₃.** Due to dynamical diffraction effects, the polarity-sensitive (A) Bragg reflections (e.g., ±(001), ±(002), and ±(003)) and (B) Kikuchi bands (e.g., ±(001)) along the polar axis exhibit intensity asymmetry. However, this intensity asymmetry can be largely affected by sample mis-tilt artifacts. Here, we simulate the CBED patterns of PbTiO₃ along [100]$_{pc}$ zone-axis with systematic crystal mis-tilt angles. (C) The intensity asymmetry in Bragg reflections can have contrast reversal with mis-tilts as small as 0.04°, while the Kikuchi bands in (D) can tolerate mis-tilt angles up to 0.23°. The diffraction patterns were simulated using 300 keV electrons with probe semi-convergence angle of 2 mrad, and PbTiO₃ thickness of 101 nm as consistent with the superlattice membrane. The intensity asymmetry is normalized as $I_{asymmetry} = (I_{hkl} - I_{\overline{hkl}})/(I_{hkl} + I_{\overline{hkl}})$.

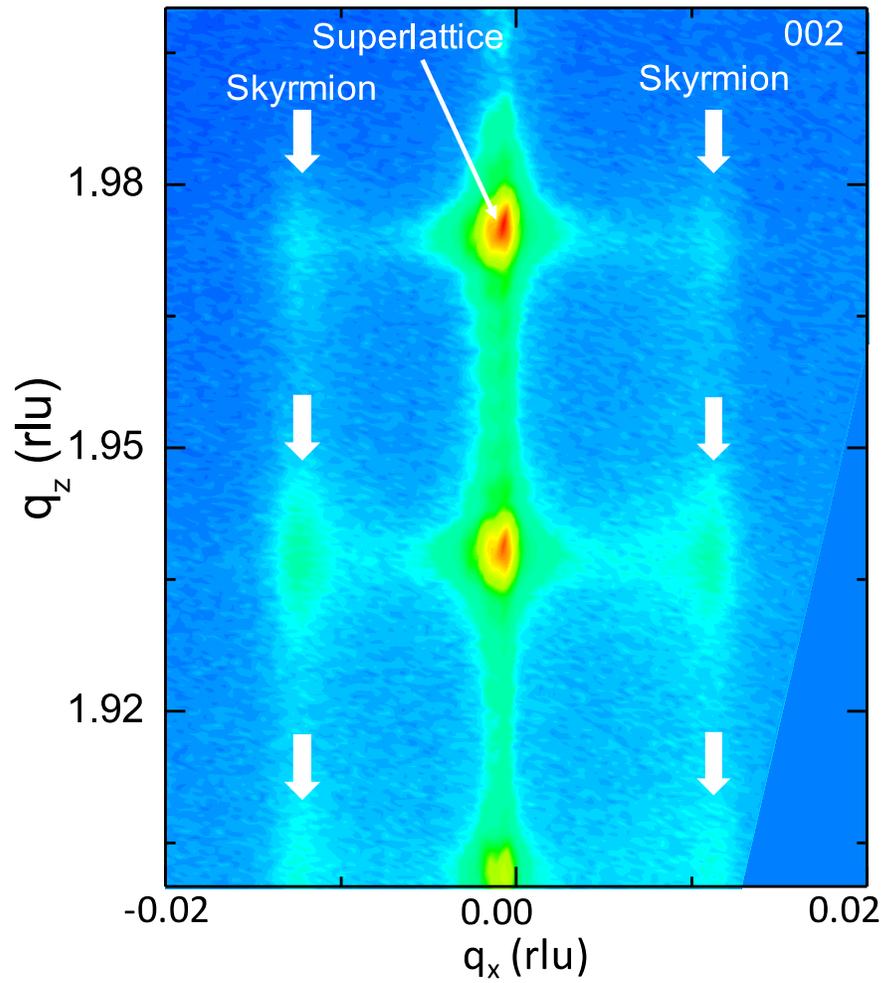

**Fig. S7. Reciprocal space mapping of the [(PbTiO$_3$)$_{16}$/(SrTiO$_3$)$_{16}$]$_8$ superlattice lifted-off membrane.** The white arrows indicate the satellite peaks associated with polar skyrmion modulation in the superlattice.

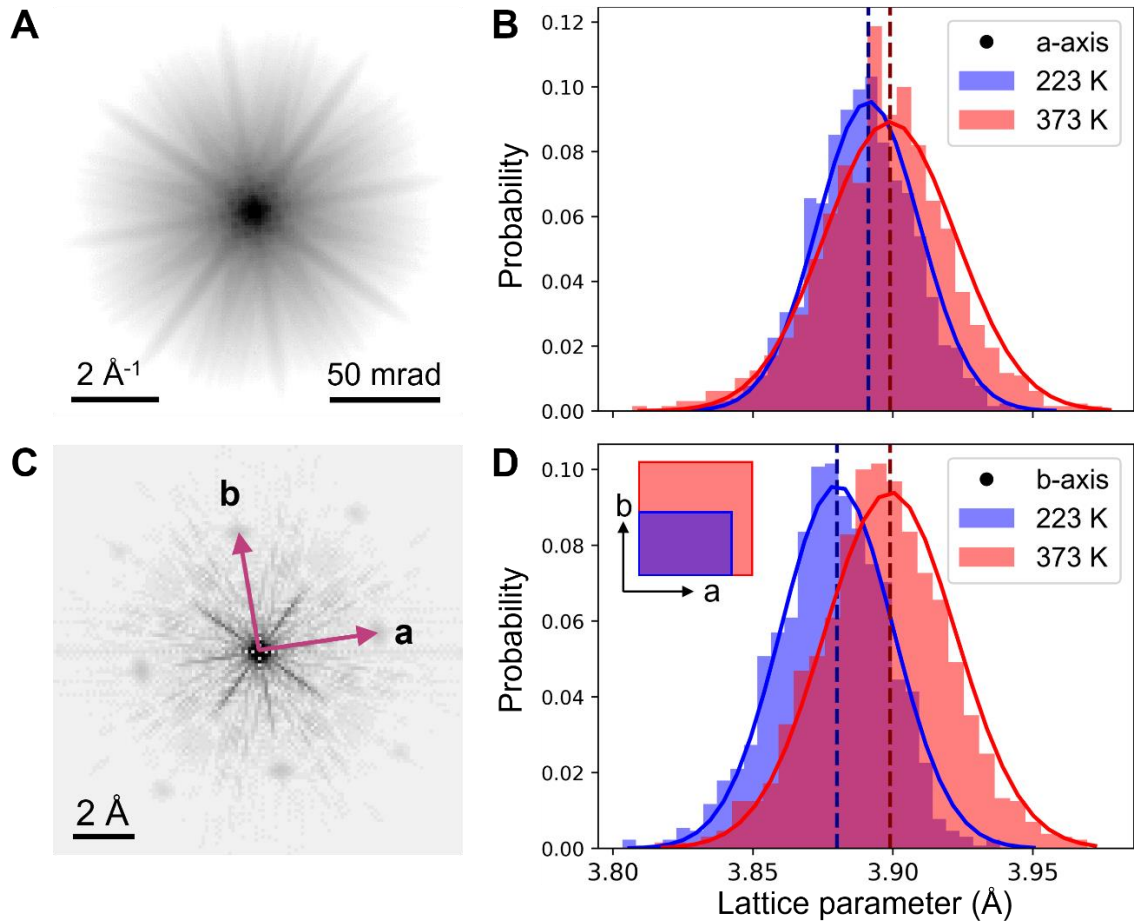

**Fig. S8. Temperature-dependent strain analysis of [(PbTiO$_3$)$_{16}$/(SrTiO$_3$)$_{16}$]$_8$ superlattice.** (**A**) Representative CBED pattern shown in logarithmic scale. (**C**) Exit wave power cepstrum (EWPC) transformation of (A) shows peaks corresponding to projected Pb-Pb inter-atomic distances in real-space. Spots that correspond to the length of projected distances along (100) and (010) are selected for tracking changes in lattice parameters *a* and *b*, respectively. Histograms of lattice parameters along (**B**) *a*- and (**D**) *b*-axis, at temperatures of 223 K (blue) and 373 K (red), over regions of ~500 nm$^2$ (>80,000 CBED patterns). The relative change in mean of *a*- and *b*-axis are ~0.2% and ~0.5%, respectively. Inset in (**D**) is an exaggerated cartoon of projected lattice parameters, indicating the film is more rectangular at 223 K than at 373 K.

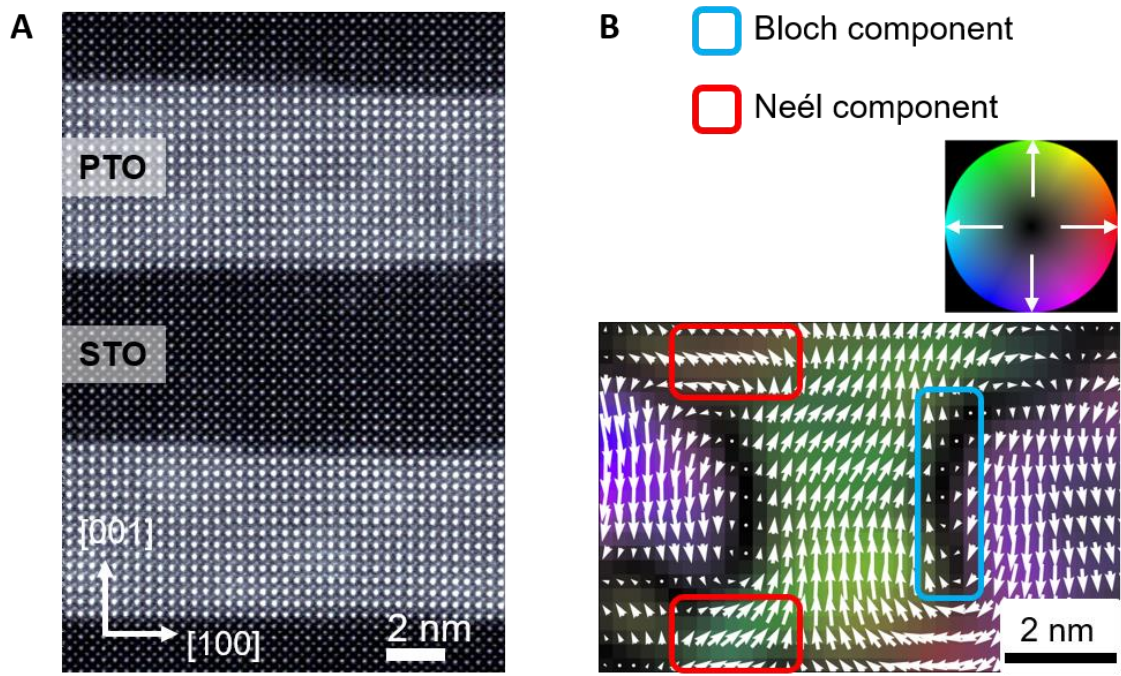

**Fig. S9. Out-of-plane polarization configurations of polar skyrmions.** (**A**) Cross-sectional HAADF-STEM image of the $(PbTiO_3)/(SrTiO_3)$ superlattice. (**B**) Polarization configuration reconstructed from the SCBED dataset, where we can access the cross-section of both Neél (red box) and Bloch (blue box) components of polar skyrmions. The out-of-plane polarizations are separated by in-plane Bloch chiral domain walls (dark regions).

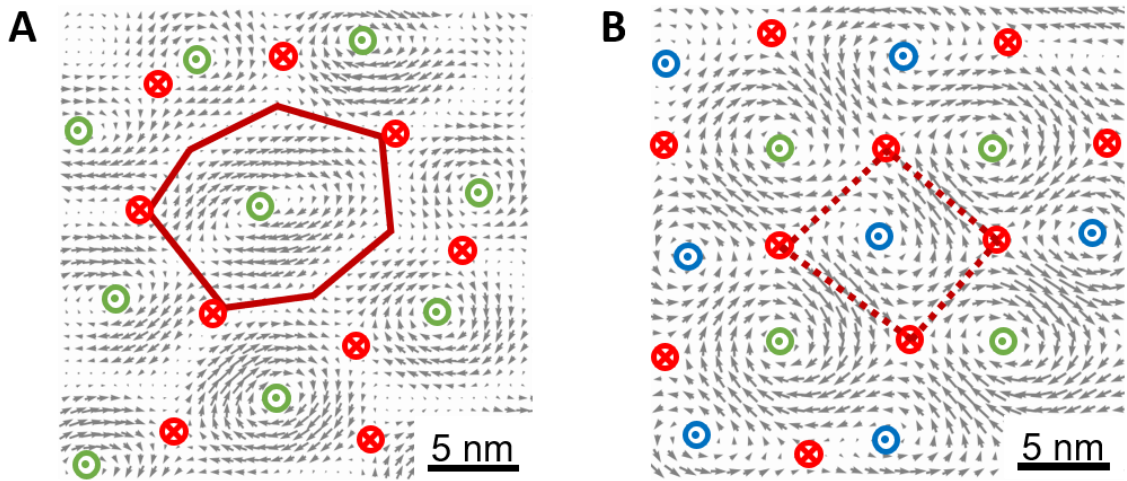

**Fig. S10. Difference in topology between a skyrmion and a meron.** Details of polarization for (**A**) a skyrmion and (**B**) a meron, where vortices (clockwise: green, counter-clockwise: blue) and anti-vortices (red) are labeled. The dots in circles represent out-of-plane polarization pointing towards out of the page, while the cross points into the page. The red solid lines in (**A**) form a closed loop of $P_{op}$ (into the page) surrounding the core of the skyrmion with an antiparallel $P_{op}$ (out of the page). In contrast, the $P_{op}$ cannot form a closed loop around the core of the meron shown in (**B**), as indicated by dashed red lines.

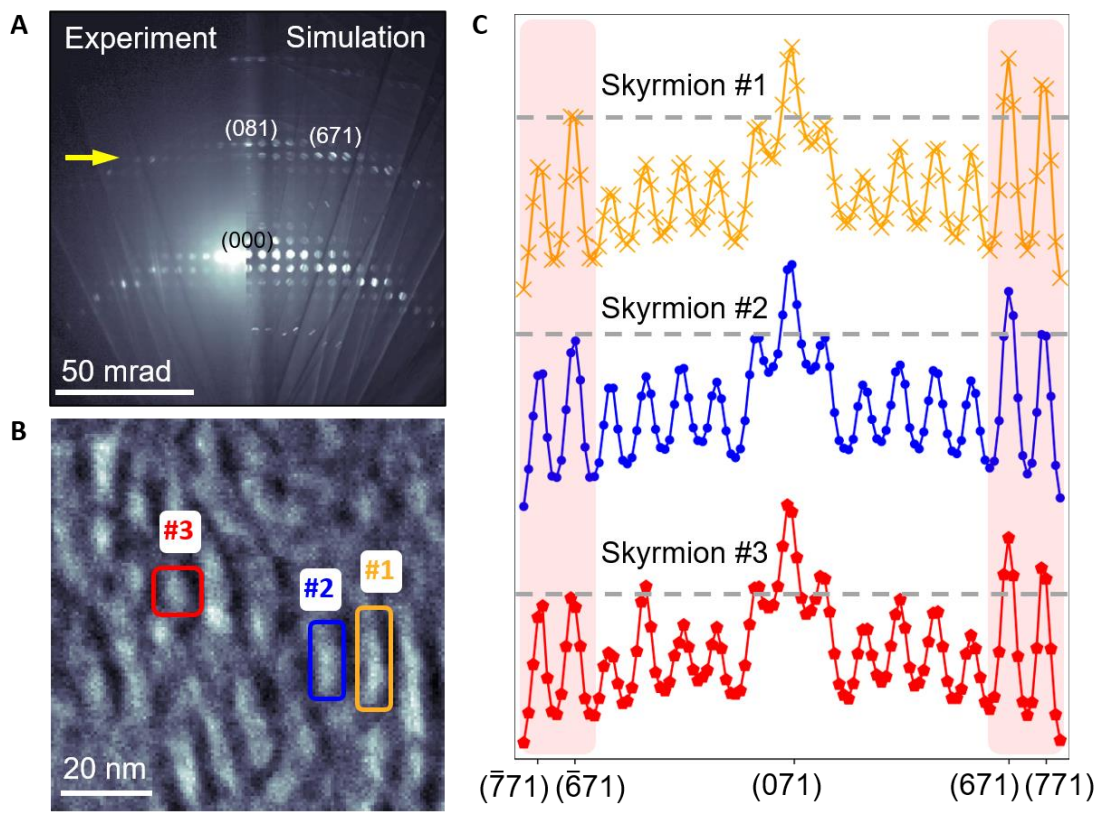

**Fig. S11. Handedness determination of the chiral polar skyrmions at room temperature.** (**A**) Representative experimental (left) and simulated (right) CBED pattern at incidence of ~6.2° away from [001] zone axis, tilted along one of the mirror planes. (**B**) Virtual dark field image reconstructed from the SCBED dataset using (081) reflection. (**C**) Intensity line profiles taken from a row of HOLZ reflections from selected regions labeled in (**B**), displaying the intensity difference between two pairs of reflections: (671) and ($\bar{6}71$), (771) and ($\bar{7}71$). The yellow arrow in (**A**) indicates the region where line profiles were drawn from.

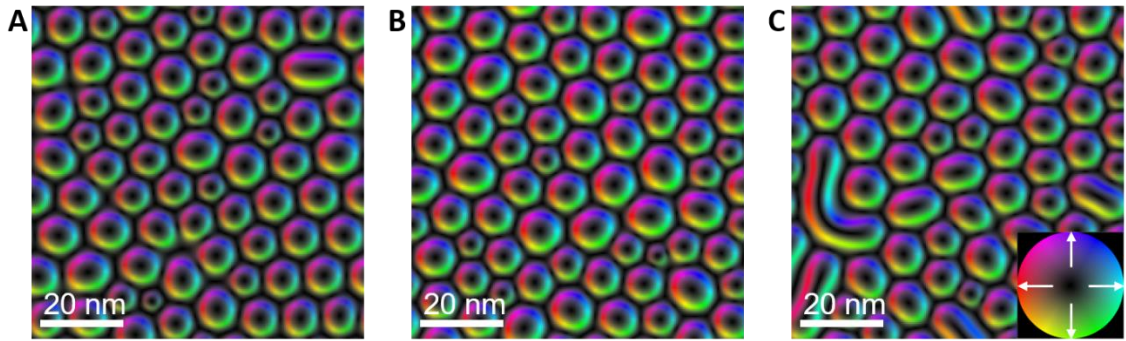

**Fig. S12. Phase field simulations of polar skyrmion textures at different temperatures and under the same strain state.** The strain is imposed on the [(PbTiO$_3$)$_{16}$/(SrTiO$_3$)$_{16}$]$_8$ superlattice by varying the substrate having in-plane lattice parameters (*a,b*) with *a=b*=3.905 Å, and at temperatures of (**A**) 223 K; (**B**) 298 K; and (**C**) 373 K. The color wheel hue (saturation) corresponds to the direction (magnitude) of the in-plane component of the ferroelectric polarization.

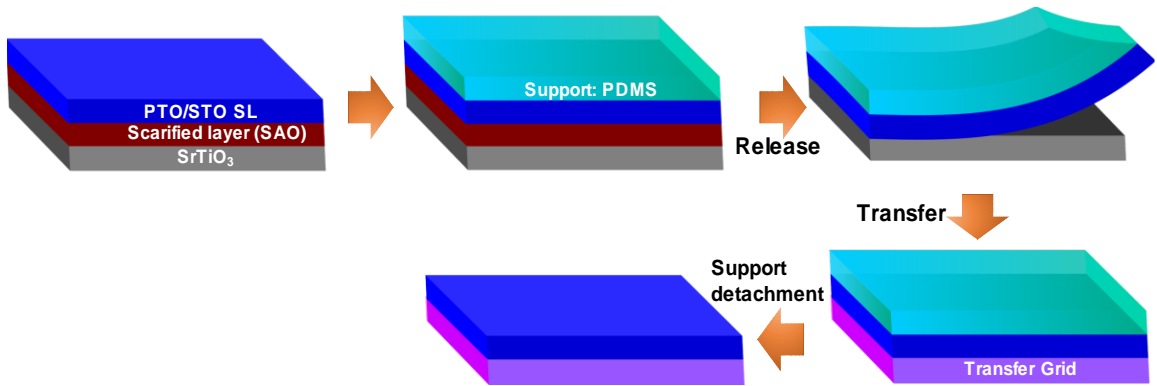

**Fig. S13. Growth and transfer of PbTiO$_3$/SrTiO$_3$ (PTO/STO) superlattice.** The Schematic of a superlattice with a Sr$_2$CaAl$_2$O$_6$ (SAO) sacrificial buffer layer. The sacrificial SAO layer is dissolved in water to release the top oxide films with the mechanical support of PMMA. The freestanding film is then transferred onto the desired substrate or TEM grid.
37

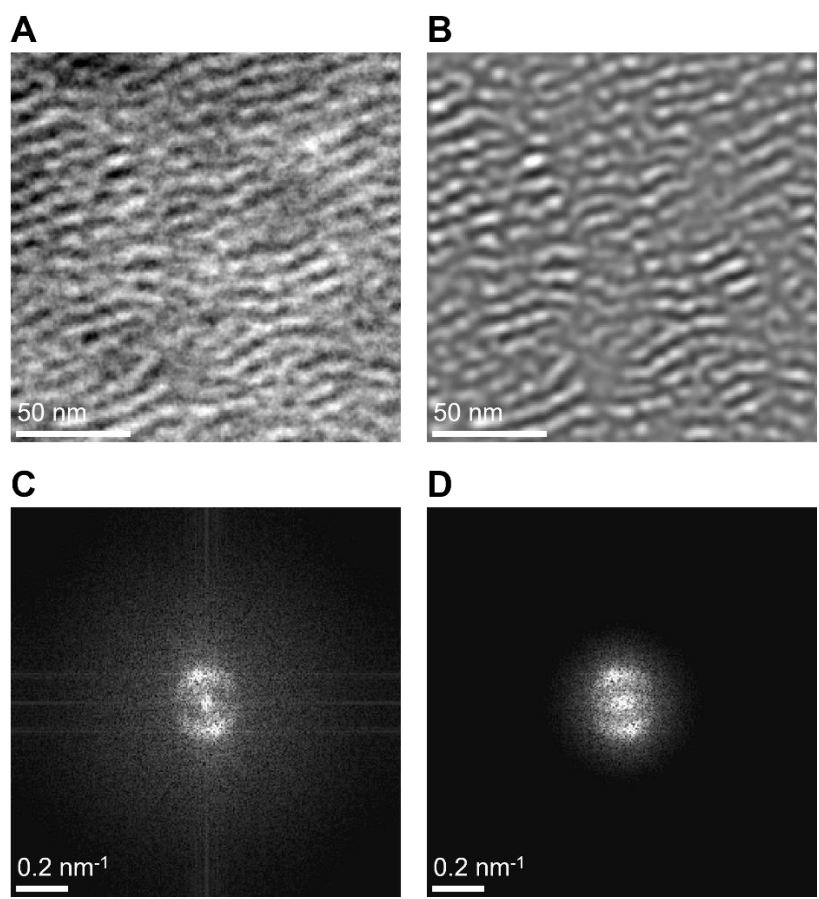

**Fig. S14. Image processing details of the polarization components image.** For clarity of display, the polarization map components reconstructed from the SCBED dataset were first subtracted off the low-frequency background with the rolling ball algorithm, then applied a Gaussian filter of 3 pixels. For example, (**A**) shows the raw $P_x$ map and (**B**) the $P_x$ after image processing. The corresponding fast Fourier transform (FFT) images of (**A**) and (**B**) were shown in (**C**) and (**D**), respectively.